\documentclass[aps,prl,twocolumn,tightenlines,superscriptaddress]{revtex4}
\usepackage{epsfig,rotating}
\usepackage{multirow}
\newcommand{\nuc}[2]{\hbox{$^{#1}$#2}}

\begin{document}
%\draft
\title{One-neutron knockout in the vicinity of the $N=32$ sub-shell closure: \nuc{9}{Be}(\nuc{57}{Cr},\nuc{56}{Cr}+$\gamma$)X}

\author{A.\ Gade}
   \affiliation{National Superconducting Cyclotron Laboratory,
      Michigan State University, East Lansing, Michigan 48824}
    \affiliation{Department of Physics and Astronomy,
      Michigan State University, East Lansing, Michigan 48824}
\author{R.\,V.\,F.\ Janssens}
    \affiliation{Physics Division, Argonne National Laboratory, Argonne,
      IL 60439}
\author{D.\ Bazin}
    \affiliation{National Superconducting Cyclotron Laboratory,
      Michigan State University, East Lansing, Michigan 48824}
\author{B.\,A.\ Brown}
    \affiliation{National Superconducting Cyclotron Laboratory,
      Michigan State University, East Lansing, Michigan 48824}
    \affiliation{Department of Physics and Astronomy,
      Michigan State University, East Lansing, Michigan 48824}
\author{C.\,M.~Campbell}
    \affiliation{National Superconducting Cyclotron Laboratory,
      Michigan State University,
      East Lansing, Michigan 48824}
    \affiliation{Department of Physics and Astronomy,
      Michigan State University, East Lansing, Michigan 48824}
\author{M.\,P.\ Carpenter}
    \affiliation{Physics Division, Argonne National Laboratory, Argonne,
      IL 60439}
\author{J.\,M.\ Cook}
    \affiliation{National Superconducting Cyclotron Laboratory,
      Michigan State University, East Lansing, Michigan 48824}
    \affiliation{Department of Physics and Astronomy,
      Michigan State University, East Lansing, Michigan 48824}
\author{A.\,N. Deacon}
    \affiliation{School of Physics and Astronomy, Schuster Laboratory,
      University of Manchester, Manchester M13 9PL, United Kingdom}
\author{D.-C.\ Dinca}
    \affiliation{National Superconducting Cyclotron Laboratory,
      Michigan State University, East Lansing, Michigan 48824}
    \affiliation{Department of Physics and Astronomy,
      Michigan State University, East Lansing, Michigan 48824}
\author{S.\,J.\ Freeman}
    \affiliation{School of Physics and Astronomy, Schuster Laboratory,
      University of Manchester, Manchester M13 9PL, United Kingdom}
\author{T.\ Glasmacher}
    \affiliation{National Superconducting Cyclotron Laboratory,
      Michigan State University, East Lansing, Michigan 48824}
    \affiliation{Department of Physics and Astronomy,
      Michigan State University, East Lansing, Michigan 48824}
\author{M.\ Horoi}
   \affiliation{Department of Physics, Central Michigan
     University, Mount Pleasant, MI 48859}
\author{B.\,P.\ Kay}
    \affiliation{School of Physics and Astronomy, Schuster Laboratory,
      University of Manchester, Manchester M13 9PL, United Kingdom}
\author{P.\,F.\ Mantica}
    \affiliation{National Superconducting Cyclotron Laboratory,
      Michigan State University,
      East Lansing, Michigan 48824}
    \affiliation{Department of Chemistry, Michigan State University,
      East Lansing, MI 48824}
\author{W.\,F.\ Mueller}
    \affiliation{National Superconducting Cyclotron Laboratory,
      Michigan State University, East Lansing, Michigan 48824}
\author{J.\,R.\ Terry}
    \affiliation{National Superconducting Cyclotron Laboratory,
      Michigan State University,
      East Lansing, Michigan 48824}
    \affiliation{Department of Physics and Astronomy,
      Michigan State University, East Lansing, Michigan 48824}
\author{J.\,A.\ Tostevin}
    \affiliation{Department of Physics, School of Electronics and
      Physical Sciences, University of Surrey, Guildford, Surrey GU2 7XH,
      United Kingdom}
\author{S.\ Zhu}
    \affiliation{Physics Division, Argonne National Laboratory, Argonne,
      IL 60439}

\date{\today}

\begin{abstract}

The one-neutron knockout reaction
\nuc{9}{Be}(\nuc{57}{Cr},\nuc{56}{Cr}+$\gamma$)X has been measured in
inverse kinematics with an intermediate-energy beam. Cross sections to
individual states in \nuc{56}{Cr} were partially untangled through the
detection of the characteristic $\gamma$-ray 
transitions in coincidence with the reaction residues. 
The experimental inclusive longitudinal momentum 
distribution and the yields to individual states
are compared to calculations that combine
spectroscopic factors from the full $fp$ shell model and nucleon-removal cross
sections computed in a few-body eikonal approach.      

\end{abstract}

\pacs{24.50.+g, 21.10.Jx, 25.60.Gc, 27.40.+z}
\keywords{\nuc{56}{Cr}, \nuc{57}{Cr}, knockout, single-particle structure}
\maketitle

Neutron-rich Ca, Ti and Cr isotopes have attracted much attention
recently. The strong proton-neutron monopole 
interaction in these exotic nuclei with $\pi f_{7/2}\nu (fp)$ configurations
causes a shift in the energy of the $\nu f_{5/2}$ neutron single-particle
orbit as protons fill the $\pi f_{7/2}$ shell and results in the
development of an $N=32$ sub-shell closure in \nuc{52}{Ca},
\nuc{54}{Ti} and \nuc{56}{Cr}~\cite{Ots01}. Nuclei in the vicinity of this new
sub-shell closure have been studied extensively in $\beta$-decay
experiments~\cite{Pri01,Man03,Man03b,Lidd},  
intermediate-energy Coulomb excitation~\cite{Din05,Bue05},
deep-inelastic heavy-ion collisions~\cite{Jan02,Forn},
%RVFJ Added a reference to "Chicken wings"
fusion-evaporation reactions~\cite{App03,Dea05,Zhu06} and secondary
fragmentation~\cite{Gad06}.  

We report on the study of the one-neutron knockout reaction
\nuc{9}{Be}(\nuc{57}{Cr},\nuc{56}{Cr}+$\gamma$)X. Direct one-nucleon
removal at intermediate beam energies~\cite{Han03} has been used
extensively to study single-particle structure in neutron-rich
and proton-rich exotic nuclear 
species (see, e.g., \cite{Nav98,Aum00,Sau00,Gad04a,Ter04,brown}).   
Single-particle cross
sections and longitudinal momentum distributions are computed using
few-body reaction theory, in the eikonal and sudden approximations
\cite{Tos99,Ber06}, and employing Skyrme Hartree-Fock
calculations~\cite{Bro98,Bro00} to reduce uncertainties in the input
to the reaction calculation. Spectroscopic factors, which relate to
the occupancy of single-particle orbits, can be deduced 
from experimental cross sections by comparison to reaction theory
and pose stringent tests for modern shell-model calculations far from
stability~\cite{Bro01}. The
longitudinal momentum distribution of the knockout residues is used --
in analogy to the angular distribution in low-energy transfer
reactions -- to 
identify the orbital angular momentum $l$ carried by the knocked-out 
nucleon.

In this paper we discuss the magnitude of the inclusive cross section, the 
small branch for the knockout to the ground state of \nuc{56}{Cr} and
the shape of the inclusive momentum distribution as compared to that
predicted on the basis of single-particle strengths from large-scale
shell-model calculations in the $fp$ shell.  

A secondary beam cocktail containing \nuc{57}{Cr} was obtained via
fast fragmentation of 
a 130~MeV/nucleon \nuc{76}{Ge} primary beam delivered by the Coupled
Cyclotron Facility of the National Superconducting Cyclotron
Laboratory at Michigan State University. The production target,
423~mg/cm$^2$ thick \nuc{9}{Be}, was located at the mid-acceptance
position of the A1900 fragment separator~\cite{a1900}. The
separator was operated at 1\% momentum acceptance.   

\begin{figure}[h]
        \epsfxsize 8.0cm
        \epsfbox{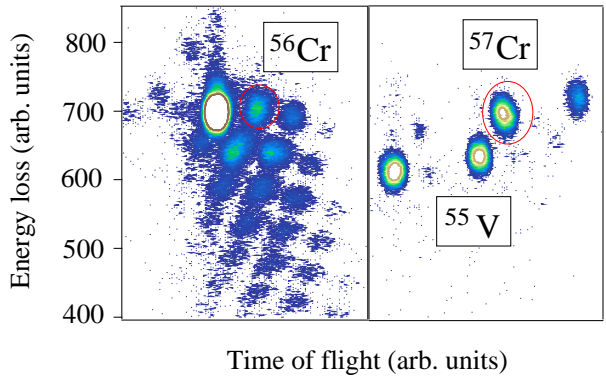}
\caption{\label{fig:pid} (Color online) Energy loss vs. time of flight
  particle identification 
  spectra for the one-neutron knockout reaction
  setting and the unreacted incoming beam. The right panel
  shows the incoming, unreacted cocktail beam passing 
  through the \nuc{9}{Be} target. This unreacted setting has been used
  to determine the normalization of the incoming rate of
  \nuc{57}{Cr}. The left panel shows the spectrum 
  obtained with the one-neutron removal reaction residues centered in
  the S800 focal plane. A software gate has been applied on incoming
  \nuc{55}{V} and \nuc{57}{Cr} in the
  time-of-flight difference measured between two plastic scintillators
  before the target.} 
%As seen in the upper panel, \nuc{55}{V} and
%  \nuc{57}{Cr} are nicely separated in energy loss but very close in
%  time of flight and thus a separation with the time-of-flight
%  difference was impossible. The lower panel shows the reaction
%  residues from both, \nuc{55}{V} and \nuc{57}{Cr}. Since one-proton pickup
%  reactions are heavily suppressed at our high beam energies, the
%  \nuc{56}{Cr} shown in the lower panel can be entirely attributed to the
%  one-neutron knockout from \nuc{57}{Cr}.} 
\end{figure}

The \nuc{57}{Cr} projectiles, with an average mid-target energy of 77
MeV/nucleon, interacted with a 375~mg/cm$^2$ thick \nuc{9}{Be} target
placed at the pivot point of the large-acceptance, high resolution
S800 magnetic spectrograph~\cite{s800}. The event-by-event particle
identification and the reconstruction of the momentum distribution of
the knockout residues were performed with the focal-plane detector
system~\cite{Yur99}. The energy loss 
in the S800 ionization chamber, the time of flight taken between plastic
scintillators and the position and angle information of the reaction 
products in the S800 focal plane were employed to
identify the reaction residues produced upon collision
with the \nuc{9}{Be} target (Fig.\ref{fig:pid}). The spectrograph was
operated in 
focus mode, where the incoming exotic beam is momentum focused
onto the secondary reaction target. The difference in the time of
flight measured between two scintillators before the secondary target
provided the particle identification of the incoming beam. Selection
of the incoming species allowed a separation of knockout residues and
fragmentation products of the different constituents of the
beam. Incoming \nuc{55}{V} and \nuc{57}{Cr} projectiles overlapped in
time of flight, but possible contamination from \nuc{56}{Cr} produced
by proton pickup on \nuc{55}{V} is highly unlikely because of momentum
mismatch. 

An inclusive cross section of $\sigma_{inc}=122(8)$~mb was determined
for the one-neutron knockout from \nuc{57}{Cr} to all bound final states of
\nuc{56}{Cr}. This was obtained from the yield of knockout residues
divided by the number  
of incoming projectiles relative to the number density of the
\nuc{9}{Be} reaction target. The fraction of incoming \nuc{57}{Cr}
projectiles in the cocktail was determined from the unreacted
spectrograph setting shown in the right panel
of Fig.~\ref{fig:pid}. The number of incoming \nuc{57}{Cr} in the
reaction setting is then derived from this fraction 
relative to scalers counting the total incoming particle flux. The
main uncertainties stem 
from the choice of the software gates used for particle identification
(2.5\%), the composition of the beam (3\%), and the momentum
acceptance of the S800 spectrograph (5\%). These systematic errors
are assumed to be independent and have been added in quadrature.

The \nuc{9}{Be} reaction target was surrounded by
SeGA, an array of 32-fold segmented HPGe detectors~\cite{sega},
arranged in a configuration with two rings (90$^\circ$ and 37$^\circ$
central angles with respect to the beam axis, equipped with ten and
seven detectors, respectively). The segmentation of the SeGA detectors
allows for an event-by-event  Doppler reconstruction where the angle of the
$\gamma$-ray emission is deduced from the position of the segment that
registered the highest energy deposition.  

\begin{figure}[h]
        \epsfxsize 8.1cm
        \epsfbox{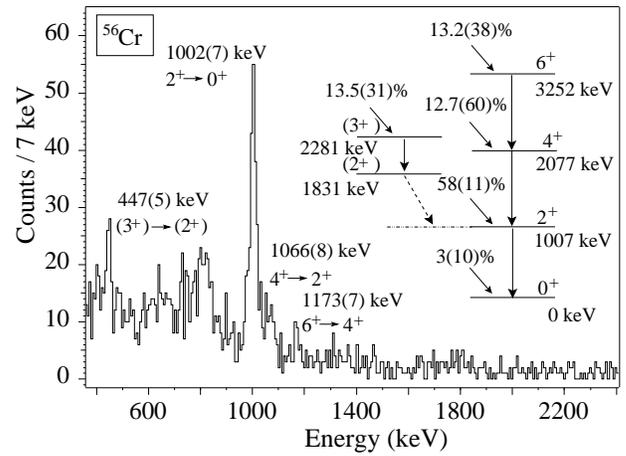}
\caption{\label{fig:spectrum} Gamma-ray spectrum detected in coincidence with
  \nuc{56}{Cr} knockout residues. The observed transitions are in
  agreement with the level scheme established in \cite{App03,Zhu06}. The
  population of individual excited states in the knockout reaction is
  derived from $\gamma$-ray intensities relative to the number of
  knockout residues taking the level scheme into account. The
  fraction of 2.6\% for the knockout to the ground state is obtained after
  subtraction.} 
\end{figure}

The SeGA total photopeak efficiency of 2.2\% for 1.33-MeV
photons energy was determined with standard
calibration sources. The latter also 
provided the detector response used to correct the in-beam data for
the Lorentz boost arising from the velocity of the emitting
reaction residues ($v/c=0.36$). 

The Doppler-reconstructed
$\gamma$-ray spectrum detected in coincidence with \nuc{56}{Cr}
reaction residues is shown in Fig.~\ref{fig:spectrum}. The
$\gamma$-ray transitions observed in the present experiment are in agreement
with the results from~\cite{App03,Zhu06} and confirm the known level
scheme, given as an inset in  Fig.~\ref{fig:spectrum}. Using the known
$\gamma$-ray branching ratios and the level scheme, the 
cross sections for the one-neutron knockout to specific final states
were deduced from the balance between the observed feeding and dexcitation
patterns. These intensity balances can be rather
uncertain in instances where unobserved $\gamma$-ray decays are
possible.

\begin{figure}[h]
        \epsfxsize 7.5cm
        \epsfbox{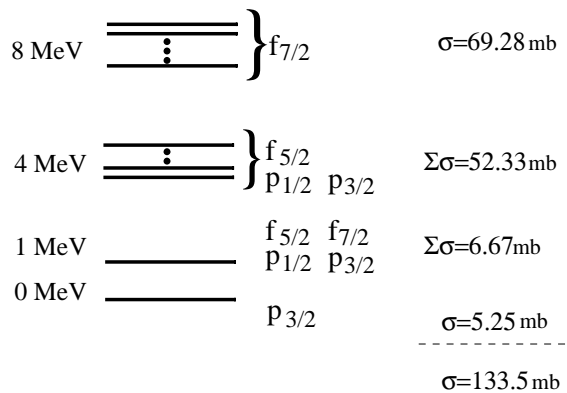}
\caption{\label{fig:sm} Schematic shell-model level scheme,
  configurations, and resulting theoretical cross
  sections for the population of excited states in \nuc{56}{Cr}. The
  shell-model calculation is schematic. The
  cross sections are calculated from eikonal reaction theory as
  described in the text (Table I).}  
\end{figure}

The spectroscopic factors leading to the ground state and the $2^+$
state (1.007~MeV) of \nuc{56}{Cr} were calculated in the full $fp$ shell-model
space with the GXPF1A interaction~\cite{Hon02,Hon04} using the codes
{\sc cmichsm}~\cite{cmichsm} and {\sc oxbash}~\cite{oxbash}.  
The results are summarized in Fig.~\ref{fig:sm} and Table~\ref{tab:theo}. The
calculations predict a $3/2^-$ ground state for \nuc{57}{Cr}, in
agreement with an early $\beta$-decay measurement~\cite{Dav78} and
consistent also with more recent experimental
observations~\cite{Man03,Dea05}. The calculations indicate a severely
fragmented single-particle strength as the derived spectroscopic factors are:  
$C^2S(0^+,p_{3/2})=0.24$, 
$C^2S(2^+,p_{3/2})=0.32$,  
$C^2S(2^+,f_{7/2})=0.03$,  
$C^2S(2^+,f_{5/2})=0.006$ and  
$C^2S(2^+,p_{1/2})=0.007$, with
the remaining spectroscopic strength, that sums up to the
neutron occupation number of 13, going to levels
up to about 8~MeV in \nuc{56}{Cr} (the neutron decay threshold is
$S_n=8.26$~MeV~\cite{AuWa}). Such high excitation energies imply that
there may be several hundred levels to consider. For an estimate of
the cross section related to these states, we take the remaining 
$(p_{1/2},f_{5/2},p_{3/2})$
strength of 
$C^2S(p_{3/2})=2.15$,
$C^2S(f_{5/2})=1.81$ and  
$C^2S(p_{1/2})=0.47$
to be centered 
at 4~MeV in excitation and the deeper $f_{7/2}$ hole strength of about 8 units
to be centered at 8~MeV in excitation energy. Clearly, the value
$C^2S(f_{7/2})=8$ should be viewed as 
an upper limit for $f_{7/2}$ occupation as it assumes that no
spectroscopic strength is lost to unbound states above the neutron
threshold. This complex scenario with spectroscopic strength to many
excited states below threshold  
leads to the following conclusion: the measured population of 
excited states quoted in Fig.~\ref{fig:spectrum} must include both the
direct population in the one-neutron knockout 
process and unobserved, discrete feeding from many higher-lying excited
states. The high partial cross section carried by the
2$^+_1$ state is likely dominated by indirect, unobserved
feeding. This state will act as doorway state funnelling the  
majority of the spectroscopic strength feeding the higher-lying, excited
states that are predicted to be populated in the one-neutron
knockout by the shell-model calculation.
      
The theoretical cross sections quoted in Fig.~\ref{fig:sm} are
obtained by combining the spectroscopic factors from the shell-model
calculations and the 
single-particle cross sections from eikonal reaction theory in the
following way: the cross sections $\sigma_i(I^{\pi})$ for the
knockout of a 
single nucleon with quantum numbers $(n,l,j)$, leaving the core in a
specific final state $I^{\pi}$, factorize into a part that 
describes nuclear structure (the spectroscopic factor $C^2S$) and a 
contribution characterizing the reaction process (the single-nucleon
removal unit cross section $\sigma_{sp}$) as

\begin{equation}
\label{eq:xsec}
\sigma_i(I^{\pi})=\sum_j \left(\frac{A}{A-1}\right)^3 C^2S(j,I^{\pi})\sigma_{sp}(j,S_n+E_f(I^{\pi}))
\end{equation}   
with summation over all the allowed angular-momentum transfers
$j$. The $A$-dependent term is a center-of-mass
correction~\cite{Die74,Bro96} to the shell-model spectroscopic factors in
the $fp$ shell. The effective separation energy of the nucleon is
$S_n+E_f(I^{\pi})$ where $S_n=5.176$~MeV is the ground-state neutron separation
energy of the projectile~\cite{AuWa} and
$E_f(I^{\pi})$ denotes the excitation energy of the final
state of the core. The single-particle cross sections are the sum of
contributions from both the stripping and diffractive 
dissociation mechanisms~\cite{Han03}:
\begin{equation}
\sigma_{sp}=\sigma_{str}+\sigma_{dif}. 
\end{equation}

The single-particle cross sections entering eq.~(\ref{eq:xsec}) were
obtained within the eikonal approach of Refs.~\cite{Han03,Tos01}.   
The stripping and diffractive contributions have been computed from
the core- and neutron-target $S$ matrices, which were calculated 
from the \nuc{9}{Be} and \nuc{56}{Cr} densities using Glauber theory,
as discussed in~\cite{Han03}. We assumed a Gaussian matter density
for \nuc{9}{Be} with a $rms$ radius of 2.36~fm. The \nuc{56}{Cr} density
was taken from SKX Hartree-Fock (HF) calculations \cite{Bro98}. The
core-neutron 
relative motion wave functions were calculated in a Woods-Saxon
potential. The diffuseness parameter was fixed at $a=0.7$~fm,
consistent with previous 
publications~\cite{Han03}. The radius parameter 
$r_0$ was selected for each single-particle orbit ($nlj$) individually to 
reproduce the $rms$ separation of neutron and core as calculated within
the SKX HF wave function. The depths of the binding potentials were
chosen to reproduce the effective separation energy of each final state. 

\begin{table}[h]
\begin{center}
 \vspace{0.5cm}
\caption{\label{tab:theo}  Results of the schematic calculation of the
  knockout process from \nuc{57}{Cr} to \nuc{56}{Cr}. Given 
  are the energy $E_f$ of the final state in \nuc{56}{Cr}, the spectroscopic
  factor $C^2S$ for the neutron knockout from orbit $nlj$ and the stripping and
  diffractive contribution to the single-particle cross section
  $\sigma_{sp}$ and resulting theoretical partial cross section $\sigma_i$ as
  calculated in the few-body eikonal reaction theory. }
\begin{ruledtabular}
\begin{tabular}{cccccccc}
$E_f$ & & $nlj$ & $C^2S$ & $\sigma_{str}$ & $\sigma_{dif}$
  &$\sigma_{sp}$ & $\sigma_i$ \\ 
(MeV) &   &    &        & (mb) & (mb) & (mb) & (mb)        \\
\hline
0.0 &$0^+_1$& $2p_{3/2}$ & 0.24 & 14.56 & 6.17 & 20.73 & 5.25 \\
1.0 &$2^+_1$&$2p_{3/2}$ & 0.32 & 12.83 & 5.15 & 17.98 & 6.07  \\
    &       &$1f_{7/2}$ & 0.03 & 9.57 & 3.18 & 12.75 & 0.40 \\
    &       &$1f_{5/2}$ & 0.006 & 8.07 & 2.61  & 10.67 & 0.07 \\
    &       &$2p_{1/2}$ & 0.007 & 12.87 & 5.17 & 18.04 & 0.13 \\ 
4.0 &    & $2p_{3/2}$ & 2.15 & 9.57 & 3.37 & 12.94 & 29.33 \\
    &    & $1f_{5/2}$ & 1.81 & 6.70 & 1.97 & 8.67 & 16.55  \\
    &    & $2p_{1/2}$ & 0.47 & 9.63 & 3.40 & 13.02 & 6.45 \\
8.0 &    & $1f_{7/2}$ & 7.97 & 6.47 & 1.78 & 8.24 & 69.28 \\
\hline
        &  &     sum     & 13  &     &       &   sum    &   133.5
\end{tabular}
\end{ruledtabular}
\end{center}
\end{table}

The large inclusive cross section for knockout from \nuc{57}{Cr}
to \nuc{56}{Cr} can be traced to the large neutron occupancy of
orbitals that, in one-neutron removal, populate bound excited states
of \nuc{56}{Cr}. Remarkably, the small theoretical cross section for 
the knockout to the ground state is also in agreement with the experimental
observation (see Fig.~\ref{fig:spectrum}) and is a consequence of the
very small 
associated spectroscopic factor of $C^2S=0.24$. In the shell model,
the $2p_{3/2}$, $1f_{5/2}$ and $2p_{1/2}$ single-particle orbits are very
close in energy, with the result that a simple description of
\nuc{57}{Cr} as a single neutron outside a $N=32$ core fails.       

\begin{figure}[h]
        \epsfxsize 7.0cm
        \epsfbox{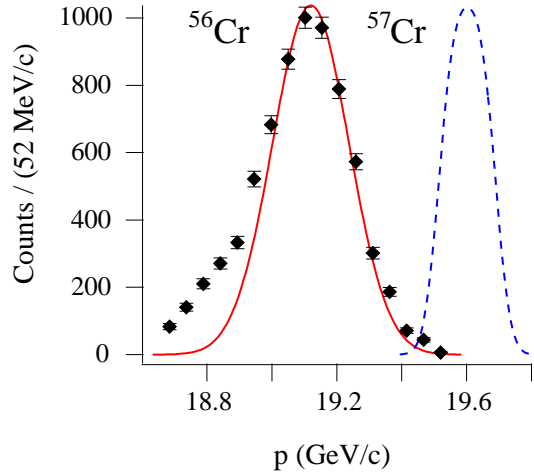}
\caption{\label{fig:mom} (Color online) Inclusive parallel momentum
  distribution of the \nuc{56}{Cr} knockout residues in comparison to
  a model calculation. The theoretical inclusive momentum distribution (solid
  red line) is calculated using the eikonal stripping reaction
  mechanism. The weights attributed to the individual contributions of
  knockout from $2p_{3/2}$, $1f_{5/2}$, $2p_{1/2}$, and $1f_{7/2}$
  orbits are taken from the corresponding theoretical cross sections
  presented in Table I. The momentum profile of the unreacted
  \nuc{57}{Cr} projectile beam passing through the target (dashed blue
  line) is also shown.}   
\end{figure}

The inclusive parallel momentum distribution of the \nuc{56}{Cr}
residues is compared with the theoretical expectation in Fig.~\ref{fig:mom}.  
The latter combines the momentum distributions for the removal of a
neutron from $2p_{3/2}$, $1f_{5/2}$, $2p_{1/2}$ and $1f_{7/2}$ orbits 
(Table~\ref{tab:theo}), calculated from the same $S$ matrices as used
for the single-particle cross sections, and the method of
\cite{Ber06,ber04}. The theoretical inclusive 
momentum distribution is then the sum of these individual distributions,
weighted with the corresponding
spectroscopic factors. The calculated shape was convoluted with the
measured momentum profile of the unreacted \nuc{57}{Cr} beam to
account for the incoming momentum spread and the straggling in the
target. The measured momentum distribution is asymmetric with the
high-momentum side being accurately reproduced by the calculation
while a tail extends toward lower momenta. Such asymmetric shapes have
been reported before~\cite{End02,Gad04b,Gad05} in nucleon knockout
reactions of well-bound systems and indicate (as yet unquantified)
effects that go beyond eikonal reaction theory. 
In the eikonal theory, the partial cross sections are determined:
(i) structurally, by the single-particle overlap for each transition,
and (ii) dynamically, by the nucleon- and residue-target S-matrices.
The expressions for the single-particle cross sections are inclusive
with respect to all final states of the target (using completeness)
when assuming these can be considered degenerate with the target
ground state for the purpose of the reaction dynamics (the adiabatic
approximation)- the result is that the $S$-matrices are calculated
at fixed energy. This is expected to be a very good approximation at
intermediate energy.
Expressing the eikonal residue momentum distribution requires additional
approximations, e.g. the neglect of residue final-state interactions, 
while this observable now probes the dynamics more closely. This
extra sensitivity indicates a small redistribution of the integrated
single-particle cross section with momentum compared to that from
the approximate (energy non-conserving) eikonal model kinematics,
with a low-energy tail characteristic of (dissipative) target
excitations that are treated only approximately (adiabatically) in
the eikonal model. The approximations required for the integrated
partial cross section on the other hand are more robust, as was
discussed above (further discussion can be found in~\cite{Gad05}).
The overall agreement in the shapes of the measured and 
and calculated inclusive momentum distributions is
consistent with the schematic distribution of single-particle
strength adopted and the shell-model
space chosen within this approach appears valid.

In summary, the one-neutron knockout reaction
\nuc{9}{Be}(\nuc{57}{Cr},\nuc{56}{Cr}+$\gamma$)X has been investigated.
The spectroscopic strength was found to be spread
over a large number of \nuc{56}{Cr} levels. Nevertheless,
calculations were able to reproduce the main observables, i.e, 
the magnitude of the inclusive cross section 
of $\sigma_{inc}=122(8)$~mb as well as the shape of the
inclusive parallel momentum distribution. These results can be viewed as
a further illustration of the potential of direct one-nucleon removal reactions
at intermediate energy and the associated theoretical framework for
spectroscopic investigations of exotic nuclei.

\begin{acknowledgments}
This work was supported by the National Science Foundation under
Grants No. PHY-0110253, PHY-9875122, and PHY-0244453, by the
U.S. Department of Energy, Office of Nuclear Physics, under 
Contract No. W31-109-ENG-38, and by the UK Engineering and Physical
Sciences Research Council grant EP/D003628.
\end{acknowledgments}


\begin{thebibliography}{10}
\bibitem{Ots01} T.\ Otsuka, {\it et al.}
%R.\ Fujimoto, Y.\ Utsuno, B.\ A.\ Brown,
% M.\ Honma, and T.\ Mizusaki, 
Phys.\ Rev.\ Lett.\ {\bf 87}, 082502
  (2001).
\bibitem{Pri01} J.I.\ Prisciandaro {\it et al.}, Phys.\ Lett.\ {\bf B510},
17 (2001).
\bibitem{Man03} P.\ F.\ Mantica {\it et al.}, Phys.\ Rev.\ C {\bf 67}, 
  014133 (2003).
\bibitem{Man03b} P.\ F.\ Mantica {\it et al.}, Phys.\ Rev.\ C {\bf
  68}, 044311 (2003).
\bibitem{Lidd} S.N.\ Liddick {\it et al}., Phys.\ Rev.\ Lett.\ {\bf 92},
   072502 (2004); Phys. Rev.  C {\bf  70}, 064303 (2004); Phys.\ Rev.\
   C {\bf 72}, 054321 (2005).
\bibitem{Din05} D.-C.\ Dinca {\it et al}., Phys.\ Rev.\ C {\bf 71},
   041302(R) (2005).
\bibitem{Bue05}  A.\ B{\"u}rger {\it et al.}, Phys.\ Lett.\ {\bf B622}, 29
  (2005). 
\bibitem{Jan02} R.V.F.~Janssens {\it et al.}, Phys.\ Lett.\ {\bf B546},
  55 (2002). 
\bibitem{Forn} B.\ Fornal {\it et al.}, Phys.\ Rev.\ C {\bf 70},
  064304 (2004); Phys.\ Rev.\ C {\bf 72}, 044315 (2005).
\bibitem{Zhu06} S.\ Zhu {\it et al.}, to be published.
\bibitem{App03} D.\ E.\ Appelbe {\it et al.}, Phys.\ Rev.\ C {\bf 67},
  034309 (2003).
\bibitem{Dea05} A.\ N.\ Deacon {\it et al.}, Phys.\ Lett.\ {\bf B622},
  151 (2005).  
\bibitem{Gad06} A.\ Gade {\it et al.}, Phys.\ Rev.\ C {\bf 73}, 037309
  (2006).  
\bibitem{Han03} P.\ G.\ Hansen and J.\ A.\ Tostevin, Annu. Rev. Nucl. Part.
Sci. {\bf 53}, 219 (2003).
\bibitem{Nav98} A.\ Navin {\it et al.},  Phys.\ Rev.\ Lett.\ {\bf 81},
  5089 (1998). 
\bibitem{Aum00} T.\ Aumann {\it et al.}, Phys.\ Rev.\ Lett.\ {\bf 84},
  35 (2000). 
\bibitem{Sau00} E.\ Sauvan {\it et al.}, Phys.\  Lett.\ {\bf B491}, 1 (2000).
\bibitem{Gad04a} A.\ Gade {\it et al.}, Phys.\ Rev.\ Lett.\
  {\bf 93}, 042501 (2004).
\bibitem{Ter04} J.\,R.\,Terry {\it et al.}, Phys.\ Rev.\ C {\bf 69},
  054306 (2004).
\bibitem{brown} B.~A.~Brown, P.\ G.\ Hansen, B.\ M.\ Sherrill, and J.\
  A.\ Tostevin, Phys. Rev. C {\bf 65}, 061601(R) (2002).
\bibitem{Tos99}
 J.\ A.\ Tostevin, J.\ Phys.\ G {\bf 25}, 735 (1999). 
\bibitem{Ber06} C.\ A.\ Bertulani and A.\ Gade, Comp. Phys. Comm., in press.
\bibitem{Bro98} B.\ A.\ Brown, Phys.\ Rev.\ C {\bf 58}, 220 (1998).
\bibitem{Bro00} B.\ A.\ Brown, W.\ A.\ Richter, and R.\ Lindsay,
  Phys.\ Lett.\ {\bf  B483}, 49 (2000).
\bibitem{Bro01} B.\ A.\ Brown, Prog.\ Part.\ Nucl.\ Phys.\ {\bf 47},
  517 (2001).
\bibitem{a1900} D.\ J.\ Morrissey {\it et al.}, Nucl.\ Instrum.\ Methods
  in Phys.\ Res.\ B {\bf 204}, 90 (2003).
\bibitem{s800} D.\ Bazin {\it et al.}, Nucl.\ Instrum.\ Methods in Phys.\
  Res.\ B {\bf 204}, 629 (2003).
\bibitem{Yur99} J.\ Yurkon {\it et al.}, Nucl.\ Instrum.\ Methods in
  Phys.\ Res.\ A {\bf 422}, 291 (1999).
\bibitem{sega}  W.\ F.\ Mueller {\it et al.}, Nucl.\ Instr.\ and
  Meth.\  A {\bf 466}, 492 (2001).
\bibitem{Die74} A.\ E.\ L.\ Dieperink and T.\ de Forest, Phys.\ Rev.\
  C {\bf 10}, 543 (1974).
\bibitem{Bro96} B.\ A.\ Brown, A.\ Csoto, and R.\ Sherr, Nucl.\ Phys.\
  {\bf A597}, 66 (1996).
\bibitem{Tos01} J.\ A.\ Tostevin, Nucl.\ Phys.\ {\bf A682}, 320c
  (2001). 
\bibitem{ber04} C.\ A.\ Bertulani and P.\ G.\ Hansen, Phys.\ Rev.\ C
 {\bf 70}, 034609 (2004).
%%%%%%%%%%%%%%%%%%%%%%%%%%%%%%%
\bibitem{Hon02} M.\ Honma, T.\ Otsuka, B.\ A.\ Brown, and T.\
  Mizusaki, Phys.\ Rev.\ C {\bf 65}, 061301 (2002).
\bibitem{Hon04} M.\ Honma {\it et al.}, in Proceedings of Fourth
  International Conference on Exotic Nuclei and Atomic Masses
  (ENAM04), Eur.\ Phys.\  J.\ A {\bf 25}, suppl. 1, 499 (2005).
\bibitem{cmichsm}
M. Horoi, B.A. Brown, and V. Zelevinsky, Phys. Rev. C 67, 034303 (2003).
\bibitem{oxbash}
B.A. Brown, A. Etchegoyen and W.D.M. Rae, The computer code OXBASH, MSU-NSCL
Report No. 524, 1998.
\bibitem{Dav78} C.\ N.\ Davids {\it et al.},
% D.\ F.\ Geesaman, S.\ L.\ Tabor, M.\
%  J.\ Murphy, E.\ B.\ Norman, and R.\ C.\ Pardo, 
Phys.\ Rev.\ C {\bf 17}, 1815 (1978).
\bibitem{AuWa} G.~Audi {\it et al.}, Nucl. Phys. A 279, 337 (2003).
\bibitem{End02} J.\ Enders {\it et al.}, Phys.\  Rev.\ C {\bf 65},
  034318 (2002).
\bibitem{Gad04b} A.\ Gade {\it et al.}, Phys.\ Rev.\ C {\bf 69},
  034311 (2004).
\bibitem{Gad05} A.\ Gade {\it et al.}, Phys.\ Rev.\ C\ {\bf 71},
  051301(R) (2005).
 

\end{thebibliography}
\end{document}